\documentclass[[final,3p,times,twocolumn    ]{elsarticle}
\usepackage{graphicx}
\usepackage{dcolumn}
\usepackage{bm}
\usepackage{amsmath}
\usepackage{comment}
\usepackage{epstopdf} 
\usepackage{xr-hyper}
\usepackage{lineno,hyperref}
\usepackage{color} 
\usepackage{soul} 

\makeatletter

\usepackage{amssymb}

\usepackage{lineno,hyperref}
\modulolinenumbers[5]

\journal{Ultramicroscopy}

\begin{document}

\begin{frontmatter}



\title{On central focusing for contrast optimization in direct electron ptychography of thick samples}

\author[emat,nanolab]{C. Gao}
\author[emat,nanolab]{C. Hofer}
\author[emat,nanolab]{T.J. Pennycook\corref{mycorrespondingauthor}}
\cortext[mycorrespondingauthor]{Corresponding author}
\ead{timothy.pennycook@uantwerp.be}

\address[emat]{EMAT, University of Antwerp, Groenenborgerlaan 171, 2020 Antwerp, Belgium}
\address[nanolab]{NANOlab Center of Excellence, University of Antwerp, Groenenborgerlaan 171, 2020 Antwerp, Belgium}

\begin{abstract}
Ptychography provides high dose efficiency images that can reveal light elements next to heavy atoms. However, despite ptychography having an otherwise single signed contrast transfer function, contrast reversals can occur when the projected potential becomes strong for both direct and iterative inversion ptychography methods. It has recently been shown that these reversals can often be counteracted in direct ptychography methods by adapting the focus. Here we provide an explanation of why  the best contrast is often found with the probe focused to the middle of the sample. The phase contribution due to defocus at each sample slice above and below the central plane in this configuration effectively cancels out, which can prevent contrast reversals when dynamical scattering effects are not overly strong. In addition we show that the convergence angle can be an important consideration for removal of contrast reversals in relatively thin samples. 
\end{abstract}




\end{frontmatter}



\section{Introduction}
The phase images produced by electron ptychography in scanning transmission electron microscopy (STEM) provide significant advantages over conventional high resolution electron microscopy, such as higher dose efficiency \cite{SSB,SSB_dose_efficiency}, post-collection aberration correction \cite{WDD_aberr_corr} and superresolution \cite{ptycho3D,Jiang_ptycho}. The wide variety of ptychographic methods all have their advantages and disadvantages, but broadly speaking we can divide them into direct \cite{SSB,WDD} and iterative methods \cite{ePIE_improved_2009,Difference_map_ptycho} of solving the phase. Although superresolution, one of the original motivations for ptychography, was first demonstrated using the direct Wigner distribution deconvolution (WDD) method \cite{Rodenburg1993304}, it is now simpler to achieve superresolution for general samples with iterative methods such as ePIE \cite{ePIE_improved_2009}. The adaption of iterative methods in electron microscopy has also been motivated by the fact that the cameras available to collect the 4D STEM data required for ptychography introduced a severe bottleneck in the speed at which the data could be collected \cite{Merlin1,pnCCD,EMPAD}. In practice this meant that extremely long dwell times had to be used, resulting in both undesirable amounts of drift and high doses using probe position samplings such as required for a typical atomic resolution STEM image. With a defocused probe and an iterative method a much lower probe position sampling can be used to obtain a high resolution image \cite{ptycho3D}, but at the cost of retaining useful signals from imaging modes requiring a focused probe, such as the Z-contrast annular dark field (ADF) signal. Practically speaking iterative methods also require longer processing times after data acquisition and greater care to obtain meaningful convergence and results. Direct methods on the other hand avoid the problems involved with iterative convergence and are fast enough to enable live imaging \cite{Live_SSB1,Live_SSB2}.

The bottleneck to scan speed imposed by cameras in 4D STEM has now been greatly reduced. Frame based cameras \cite{Medipix3,pnCCD_2016,EMPAD,DECTRIS} continue to increase in speed and event driven cameras \cite{Timepix3} have now completely removed the speed bottleneck compared to the most rapid of conventional STEM imaging. Focused probe ptychography can now be performed with very little drift and very low doses indeed, and the ADF workflow that has made STEM so popular in materials science can now be complemented with simultaneously acquired ptychography. A common criticism of the direct ptychography methods has been the approximations used in the theory motivating them. The WDD method invokes the multiplicative approximation \cite{WDD_aberr_corr,WDD} and the single side band (SSB) \cite{SSB,SSB2} method invokes the weak phase approximation. However the fact that these approximations are used to motivate the methods does not necessarily mean the methods are not useful well beyond the validity of these approximations, and various experiments now show that they can indeed provide very valuable information about samples that are of thicknesses typical for materials science. This includes clearly locating the light elements that are generally hidden by nearby heavy columns in conventional STEM imaging \cite{WDD,Chuang_APL,Laura}. 

However the contrast of the ptychographic images is not always without complication as contrast reversals can appear with both direct and iterative methods, often with the columns taking on what can be described as a donut or caldera shape \cite{WDD,Chuang_APL,Laura}. Such contrast reversals have also been observed in integrated Center of Mass (iCoM) images \cite{Laura}. Recent investigations have revealed that for direct ptychography, as in iCoM, optimal contrast often occurs with the probe focused to the center of the sample for quite a range of thicknesses \cite{Chuang_APL,Laura}. Here we explain how the equal and opposite phase contributions of the defocus at sample slices above and below the central slice effectively cancels out when the probe is focused to the central slice of the sample. This explains why focusing to the middle of a sample often provides the best contrast, at least in samples in which strong dynamical scattering does not overly unbalance the contributions from above and below the central plane. We also show how the convergence angle can be an important consideration for avoiding contrast reversals, particularly in thinner samples with significant projected potentials.


\section{Theory}

In both SSB and WDD ptychography, the intensity captured by the camera is Fourier transformed with respect to probe position resulting in
\begin{equation}\label{G pattern}
G(\mathbf{K}_f,\mathbf{Q}_p)=A(\mathbf{K}_f)A^*(\mathbf{K}_f+\mathbf{Q}_p)\otimes_{\mathbf{K}_f}\Psi_s(\mathbf{K}_f)\Psi_s^*(\mathbf{K}_f-\mathbf{Q}_p),
\end{equation}
where $Q_p$ is the spatial frequency, $\Psi_s$ denotes the Fourier transform of the specimen transmission function, and $A(\mathbf{K}_f)=a(\mathbf{K}_f)\exp[i\chi(\mathbf{K}_f))]$ is an aperture function describing both the extent of the aperture via the top hat function $a(\mathbf{K}_f)$ and any lens aberrations present via the aberration function $\chi(\mathbf{K}_f)$. To motivate the SSB ptychography method, the WPO approximation is applied to simplify the expression to 
\begin{eqnarray}\label{ssb_Eq}
G(\mathbf{K}_f,\mathbf{Q}_p)&=&|A(\mathbf{K}_f)|^2\delta(\mathbf{Q}_p)\nonumber\\&&+A(\mathbf{K}_f)A^*(\mathbf{K}_f+\mathbf{Q}_p)\Psi_s^*(-\mathbf{Q}_p)\nonumber\\&
&+A^*(\mathbf{K}_f)A(\mathbf{K}_f-\mathbf{Q}_p)\Psi_s(\mathbf{Q}_p),
\end{eqnarray}
from which the object transmission function $\Psi_s(Q_p)$ at spatial frequency $Q_p$ can be obtained by integrating the phase and amplitude contained in regions passed by both $A^*(K_f)$ and $A(K_f+Q_p)$, excluding those regions in which $\Psi^*_s(-Q_p)$, is also passed by $A^*(K_f+Q_p)$ and destructive interference occurs due to $\Psi^*_s(-Q_p)$ being out of phase with $\Psi_s(Q_p)$. The contrast transfer function (CTF) is therefore linked to the double A functions, $A^*(K_f)A(K_f-Q_p)$, or equivalently in the $A(K_f)A(K_f+Q_p)$ by the same argument, via the frequency dependent size of the double overlap regions \cite{SSB2, Column}.

\begin{figure}[!ht]
\center\includegraphics[width=0.38\textwidth]{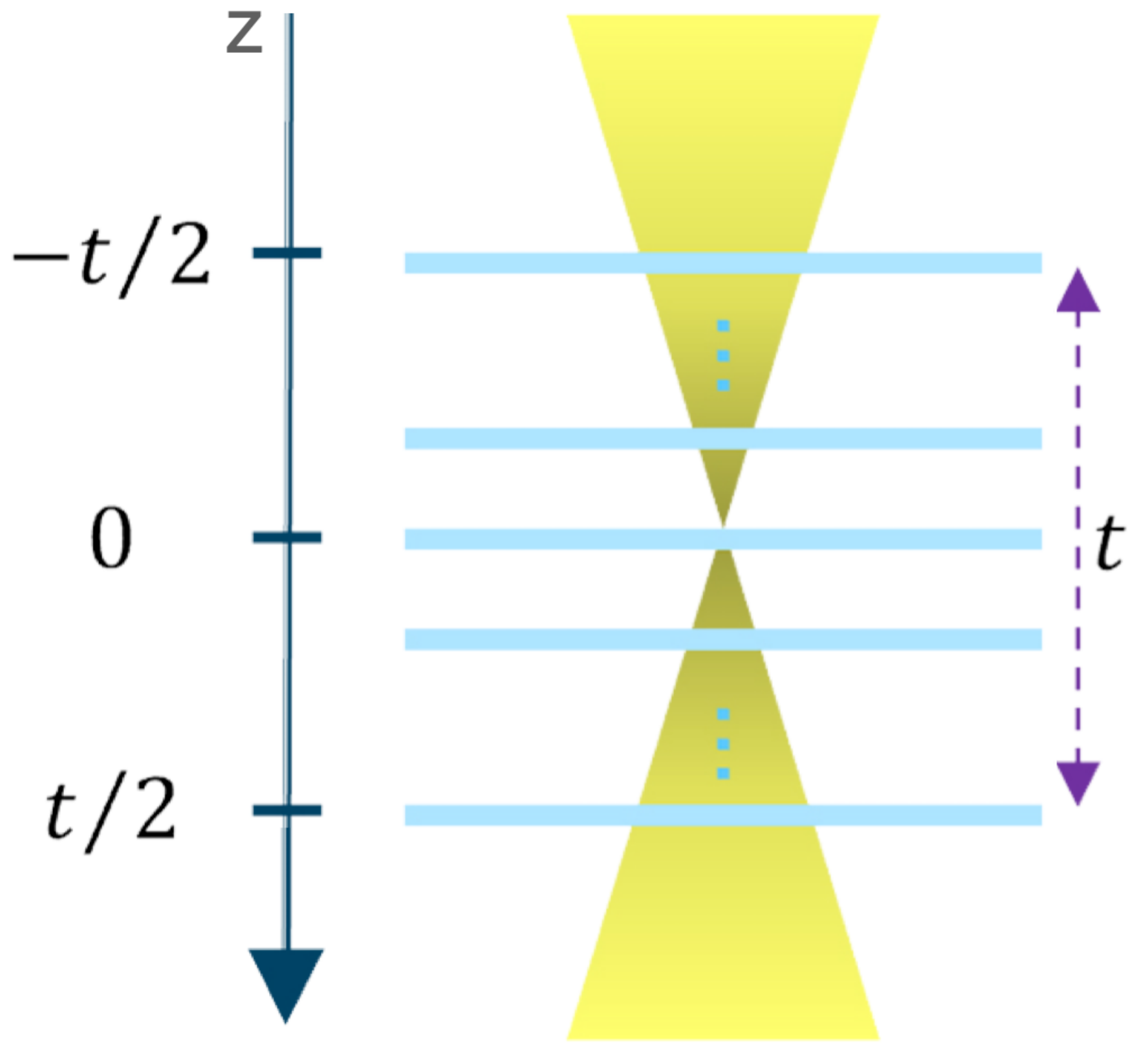}%
 \caption{\label{slices} {Illustration of how different regions of the probe (yellow) interact with different slices of the sample (blue) depending on the thickness ($t$) of the sample. The probe defocus is therefore different for the different slices of the sample. In this case the probe is focused to the center of the sample, but in general the defocus at each slice of sample depends on both the thickness of the sample and the focal point of the probe.} }
\end{figure}
In the absence of aberrations these double A functions, the multiplied pairs of aperture functions, are purely real, and the phase measured is entirely due to the specimen transmission function. However defocus is only zero at the focal point of the probe and if the sample is many nanometers thick, as is common, the probe will not be focused throughout the sample, as illustrated in figure \ref{slices}. The defocus will then vary significantly throughout the thickness of the sample introducing phase that is not purely from the object function. If we take the $z$-axis as corresponding to that along which the probe propagates and explicitly insert $z$-dependent defocus into one of the double A functions we obtain, in the absence of other aberrations,
\begin{equation}\label{doubleA}
A^*(\mathbf{K}_f, z)A(\mathbf{K}_f-\mathbf{Q}_p, z)=\\a(\mathbf{K}_f)a(\mathbf{K}_f-\mathbf{Q}_p)e^{i\pi\lambda \xi\Delta z },
\end{equation}
where 
\begin{equation}
\xi=|\mathbf{K}_f-\mathbf{Q}_p|^2 -|\mathbf{K}_f|^2=|\mathbf{Q}|^2_p-2\mathbf{K}_f\cdot\mathbf{Q}_p,
\end{equation}
and $\Delta z$ is the z-distance from the focal plane of the probe. 

We now adapt the concept of the integrated CTF (iCTF) introduced for ABF  \cite{ABF_iCTF} for SSB ptychography. For a sample with thickness of $t$, if we take the middle plane of the sample as the reference plane for the probe, the zero defocus point, the range of defocus value is from $-t/2$ to $t/2$, where positive values denote underfocus and negative values denote overfocus. If instead the probe is defocused from the central plane an amount $\Delta f$, then the integrated double A function
\begin{equation}
    \begin{array}{l}
     \mathrm{iAA}=   \frac{1}{t}\int_{\Delta f-t/2}^{\Delta f+t/2}\, A^*(\mathbf{K}_f,z)A(\mathbf{K}_f-\mathbf{Q}_p,z)\,dz
    \end{array}  
\end{equation}
or
\begin{equation}
    \begin{array}{l}
  \mathrm{iAA}=\frac{a(\mathbf{K}_f)a(\mathbf{K}_f-\mathbf{Q}_p)}{i\pi\lambda \xi t} \left[ e^{i\pi\lambda \xi(\Delta f+t/2)}-e^{i\pi\lambda \xi(\Delta f-t/2)}\right]
    \end{array}  
\end{equation}
Note that the iAA remains a function of $\mathbf{Q}_p$, $\mathbf{K}_f$ and the thickness $t$, and is a complex object retaining phase due to defocus. The \textit{complex} iCTF is then the same as the iAA with the triple overlapped area excluded. 

\section{Results and Discussion }
\begin{figure}[!ht]
 \center\includegraphics[width=\columnwidth]{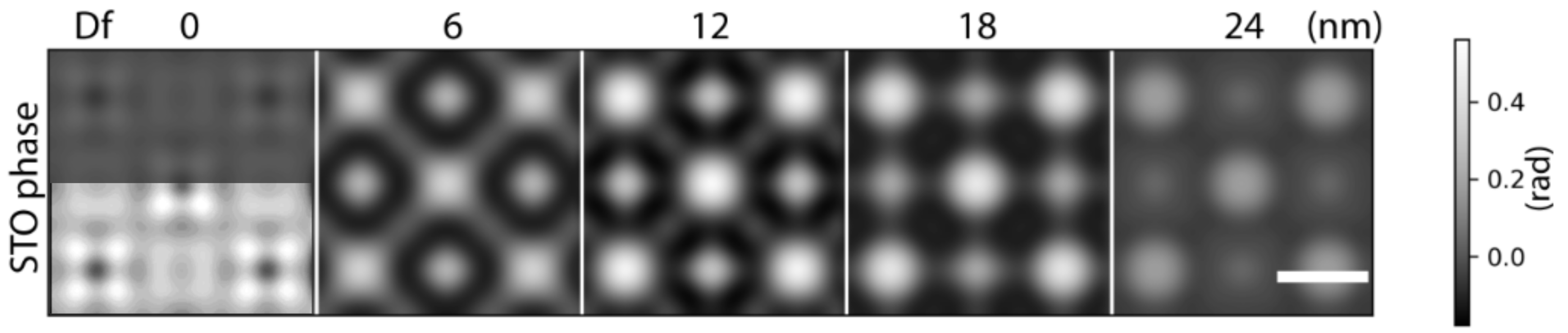}%
 \caption{\label{24nmsto} {SSB ptychography images of 24 nm thick STO simulated noise free with the probe focused to different depths of the sample from the entrance surface (0 nm of defocus) to the exit surface (24 nm of defocus). Overlaid on the bottom of the zero defocus image is a contrast boosted version to facilitate seeing the contrast reversals. The scale bar is 2 $\AA$.} }
 \end{figure}

 Figure \ref{24nmsto} shows noise free SSB ptychography images of 24 nm thick STO simulated using the multislice algorithm with the probe focused to different distances from the entrance surface. The simulations were performed using abTEM \cite{abtem} in combination with pyPtychoSTEM \cite{pyPtychoSTEM}. A 200 kV accelerating voltage, a 20 mrad convergence angle, and a scan step size of 0.2 \AA~were used. The entrance surface of the sample is generally the optimal focal plane for ADF imaging, but this focus produces contrast reversals in every atomic column of the STO in the SSB image with this thickness of STO. Furthermore, the overall contrast in the SSB image is much lower compared to focusing the probe more into the middle of the sample. 
Consistent with our previous simulations of other STO thicknesses up to 28 nm \cite{Chuang_APL}, the optimum contrast occurs with the focal point of the probe midway through the sample. Furthermore, from the images shown using the same grayscale rendering, this central focal point clearly produces both contrast reversal free images and the strongest contrast overall. 

\begin{figure}[!ht]
 \center\includegraphics[width=0.48\textwidth]{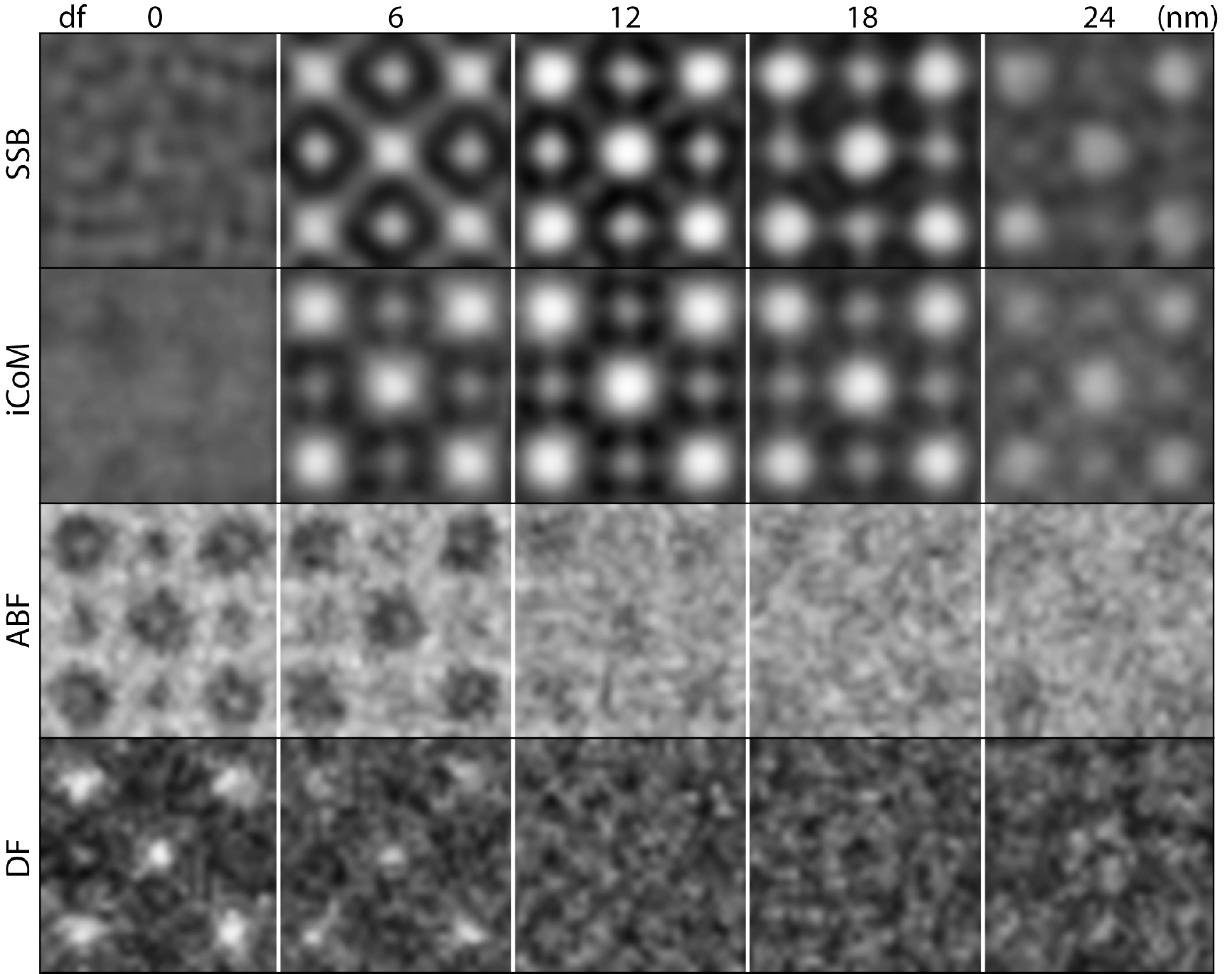}%
 \caption{\label{DF_ABF_icom_SSB} {Probe focal point dependence (with zero defocus (df) defined here as the entrance surface of the sample) of the contrast of the SSB, iCoM, ABF, and ADF signals simulated at a dose of 2000 e$^-/$\AA $^2$ for 24 nm thick STO. 
 } }
 \end{figure}

Figure \ref{DF_ABF_icom_SSB} compares SSB, iCoM, ABF and ADF images simulated for the same STO projection as figure \ref{24nmsto}, again as a function of how far within the sample the probe is focused, but now with the finite dose of 2000 e$^-/$\AA $^2$.
The ABF, iCoM and SSB ptychography are all sensitive to the light atoms, with the iCoM and SSB signals capable of providing far stronger signal than the ABF. However it is only with the probe focused more towards the center of the sample that iCoM and SSB provide the strongest contrast. With the probe focused to the entrance surface, with this particular sample configuration, the ABF actually provides stronger contrast. It can also be seen here how the ADF contrast reduces as the focus is moved further into the sample from the entrance surface. While this focal dependence has been explained for ABF \cite{ABF_iCTF} and iCoM \cite{iCoM}, so far relatively little has been published on this for direct ptychography beyond observation \cite{Chuang_APL,Laura}. 

\begin{figure}[!ht]
\center\includegraphics[width=0.48\textwidth]{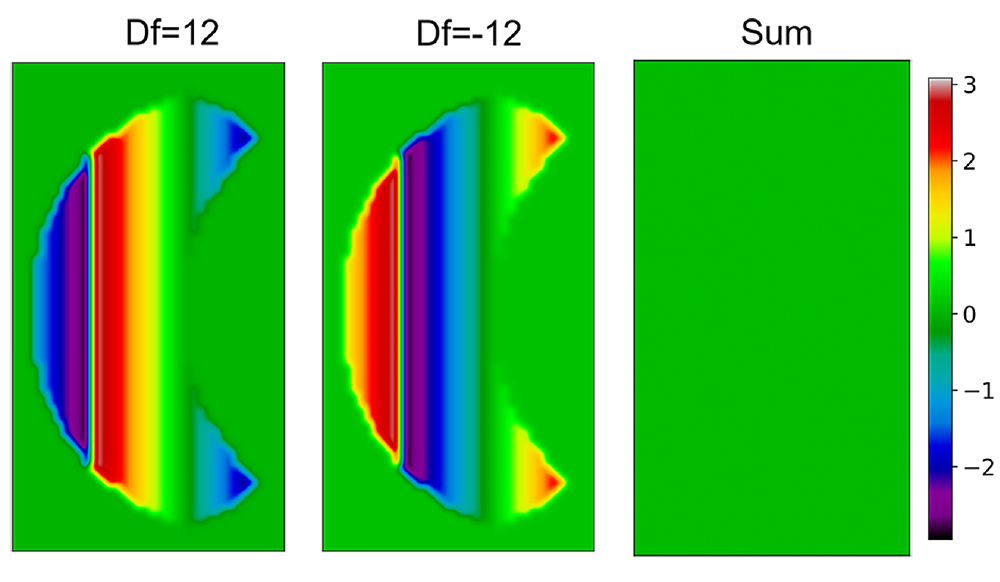}%
 \caption{\label{trotters1} {Defocus contribution to the phase in the double disk overlap region of a single spatial frequency at $\pm12$ nm of defocus. The defocus phase contribution in the two slices is perfectly out of phase, thus when they are summed the resulting phase is zero. } }
\end{figure}
In the absence of other aberrations, at the plane at which the probe is focused there is no defocus contribution and phase only comes from the object function. For a thin weak object the phase is flat across the double disk overlap regions. However defocus  induces phase variation within the double disk overlaps such as that illustrated in figure \ref{trotters1}. As sample thickness increases so too does the amount of defocused probe interacting with the sample, as shown schematically in figure \ref{slices} and mathematically in equation \ref{doubleA}. For a focused probe configuration this means a much increased amount of defocus contributing to direct ptychography images from sample planes away from the focal plane of the probe. From equation \ref{doubleA} it can also be appreciated that how the defocus weights the different depths of the sample depends on how the probe is focused relative to the sample, as this will change the defocus at each plane within the sample. 

\begin{figure}[!ht]
\center\includegraphics[width=0.48\textwidth]{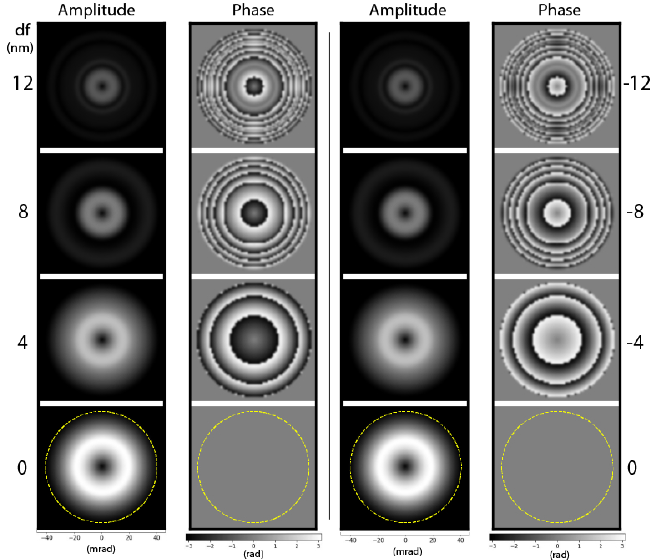}%
\caption{\label{AAslices} {Overall phase and amplitude from defocus vs spatial frequency at different distances from the focal plane of the probe (df$ = 0$). As can be seen, the contributions from planes equally above and below the focus of the probe contribute with equal magnitude but opposite phase. Note there is no object function here, and the lack of any phase at the central plane means there will be no defocus contribution to the phase here and only the phase due to an object function would contribute here to the SSB image.} }
\end{figure}

\begin{figure}[!ht]
\center\includegraphics[width=0.48\textwidth]{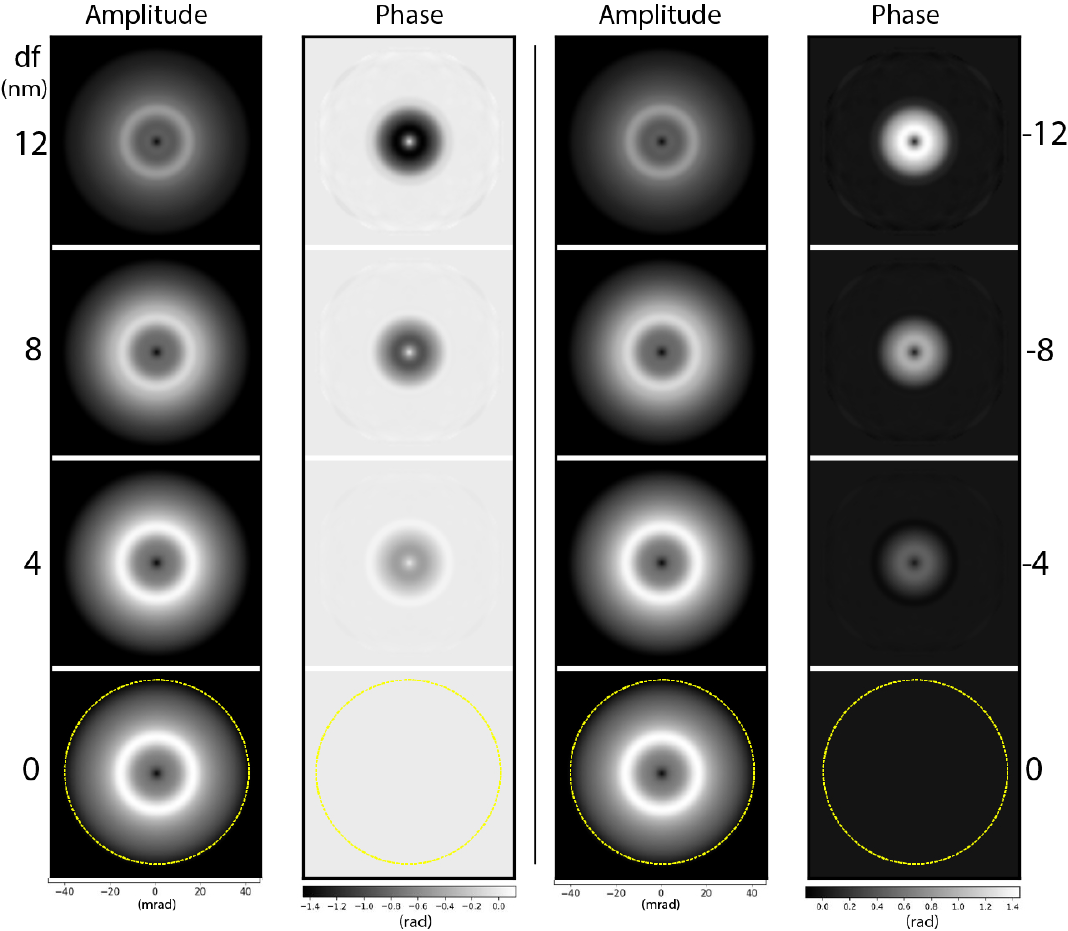}%
 \caption{\label{Integrated} { Integrated defocus induced phase and amplitude plotted vs spatial frequency with the probe focused to different points within the 24 nm integration interval. Here defocus (df) indicates the distance from the center of that 24 nm region. Note again there is no object function here, so the lack of phase shown here when the probe is focused to the central plane (df$ = 0$) means the object function phase is transferred without a phase contribution from defocus.}}
\end{figure}

To  illustrate the effect of the defocus as a function of the distance from the focal point of the probe we plot the phase and amplitude of a double A function vs spatial frequency in Figure \ref{AAslices} for defocus values from -12 to 12 nm for the same conditions used previously. This corresponds to the range of defocus values that contribute to the SSB image of the 24 nm thick STO sample with the probe focused to the central plane. Positive defocus values denote underfocus and negative values denote overfocus. Although the complexity of the phases increases with the magnitude of the defocus, it can be seen that equal and opposite values of defocus induce equal and opposite phases over the whole range of spatial frequencies. The amplitudes shown in the figure include the exclusion of the triple overlap regions as required in the SSB method, and thus reflect both the size of the pure double overlap regions in probe reciprocal space at each spatial frequency as well as the influence of the defocus. 
When defocus induced phase variations inside the double disk overlaps are integrated destructive interference occurs, diminishing the amplitude of the spatial frequencies as seen in figure \ref{AAslices} for increasing levels of defocus. This is the effect that enables optical sectioning \cite{WDD_aberr_corr}, but it must be noted that the contribution of slices away from focus is not zero even if diminished.

Now consider two slices of sample with the same object function, equal amounts above and below the central plane of the sample. With the probe focused to the central plane of the sample, the defocus of the two slices is equal and opposite. Therefore, if we neglect multiple scattering, and the sample is the same in these two slices, the contribution of defocus on these two planes will cancel out. This is illustrated for a single spatial frequency in figure \ref{trotters1}. 

If we consider the effect on the electron beam as the coherent summation of the effect of each slice of the sample, as in the iCTF concept used to explore the optimal focus of ABF imaging \cite{ABF_iCTF}, when the probe is focused to the central plane of the sample, the defocus phase contribution to every other plane has a matching plane in which the defocus is exactly the opposite. Thus the cancellation of the phases illustrated in figure \ref{trotters1} occurs for each spatial frequency through the whole thickness of the sample, cancelling out the defocus phase contributions to the final phases reaching the detector. Additionally, the optical sectioning effect means that total contribution of planes away from the focal point of the probe are diminished. 

When the probe is not focused to the central plane of the sample, not all planes of the sample will have matching planes with the opposite value of defocus, and there will remain a significant defocus phase contribution to the final image. We can illustrate this using the iCTF concept adapted for SSB ptychography and integrate the double A function over the sample thickness to observe the cumulative effect of defocus over all the slices in a sample on the total extracted phase and amplitude. The iCTF for SSB ptychography is derived as shown in the supplementary information. The results are shown in Figure \ref{Integrated} for the probe focused from the entrance surface though to the exit surface for the 24 nm thick STO. Keeping in mind no influence of the object function is included in these plots, when the probe is focused to the central plane of the sample, the defocus contributions to the phase from slices above and below the central plane cancel out resulting in a net zero phase in the double A functions, as seen in the figure. When the probe is instead focused away from the center of the sample the defocus contribution becomes increasingly asymmetric across the sample thickness with increasing distance from the central sample plane, resulting in significant phase influence from the defocus itself in the final image. 

These effects can explain why the optimal focal plane of the probe for direct ptychography is often in the centre of the sample for intermediate thickness samples. When there exist slices with a defocus not balanced with a slice elsewhere in the sample, the defocus effect on the phase contribution of these slices is significant. The greater the amount of sample thickness without identical slices of sample at an equal and opposite defocus, the greater the overall contribution of defocus phase to the final phase image. When the probe is centered exactly at the central plane the defocus contributions cancel at every plane of the sample away from the central slice, again assuming the sample is uniform. If the sample is not uniform then the interaction with the defocused probe will alter somewhat even if the overall effect is likely often similar in terms of optimal focus being relatively near the central plan of the sample.  

We emphasise that this analysis of the defocus influence across the thickness of the sample and the cancellation of the defocus phase in the slices above and below the probe focus ignores the effects of dynamical scattering. Dynamical effects will be sample dependent, and it is interesting to note that while with STO central focusing has been observed to be optimal for 16, 20, 24 and 28 nm \cite{Chuang_APL}, Clark et al. observed a band of thicknesses for GaN in which central focusing did not remove the caldera shaped contrast reversals \cite{Laura}. It seems noteworthy that this band of thickness was within the range one would consider quite thin samples and are within the 17 nm depth of field for the 14.4 mrad convergence angle and 300 kV conditions they used. This is in contrast to the results here and all the thick samples examined in Gao et al., where the samples are all thicker than the depth of field.

\begin{figure}[!ht]
\center\includegraphics[width=0.48\textwidth]{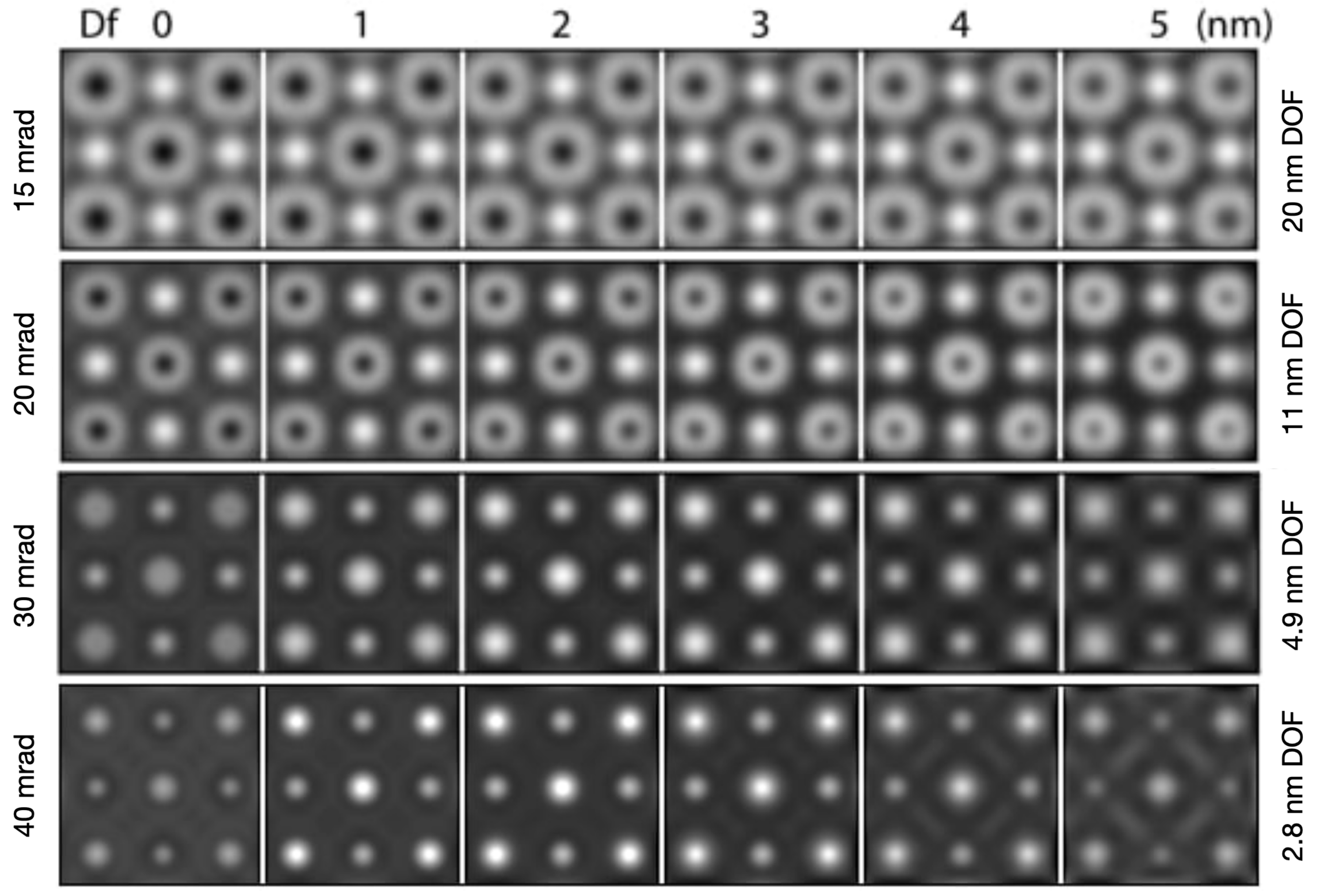}%
 \caption{\label{5nm_vsca} { SSB images of very thin 5 nm thick STO as a function of physical probe focus within the sample depth (zero defocus (df) is again the entrance surface) as a function of the convergence angle. As the convergence angle increases such that depth of field is on par or shorter than the thickness the contrast reversals are overall diminished or absent and central focusing again provides the best contrast.}}
\end{figure}

We therefore now test what happens with much thinner 5 nm thick STO as a function of both physical probe focus and convergence angle in figure \ref{5nm_vsca}. For reference Ga is atomic number 31 while Sr is 38 and Ti 22, and on the lighter side N is atomic number 7 while O is 8. This thickness of STO is therefore similar to the range of thickness of GaN which Clark et al.~found continued to exhibit caldera with central focusing. We also find that central focusing makes relatively little difference with such a thin sample, when smaller convergence angles are used such that the depth of field is significantly larger than the sample thickness. Donuts appear on the heavier Sr and Ti columns regardless of which plane within the sample is focused exactly. However, when the convergence angle is increased the donuts close up and we again find central focusing becomes optimal. 

At the 200 kV we used, a 30 mrad convergence angle puts the depth of field at just under the 5 nm thickness and at 40 mrad the depth of field is significantly less than the thickness of the sample. At 30 mrad there remains a very slight dip in the heaviest columns with the probe at the entrance surface. At 40 mrad this dip has disappeared, but for both 30 mrad and 40 mrad the strongest contrast is still seen with the probe focused to the center of the sample.

With the whole sample in focus the effect of defocus within the sample seems natural to make less difference than when the sample thickness exceeds the depth of field. Even though the defocus values are not by themselves different when the probe is focused to different points within the sample with different convergence angles, the effect on the double disk overlaps as a whole should be reduced with the sample entirely within the depth of field, as can be understood in terms of the optical sectioning effect.

\begin{figure}[!ht]
\center\includegraphics[width=0.48\textwidth]{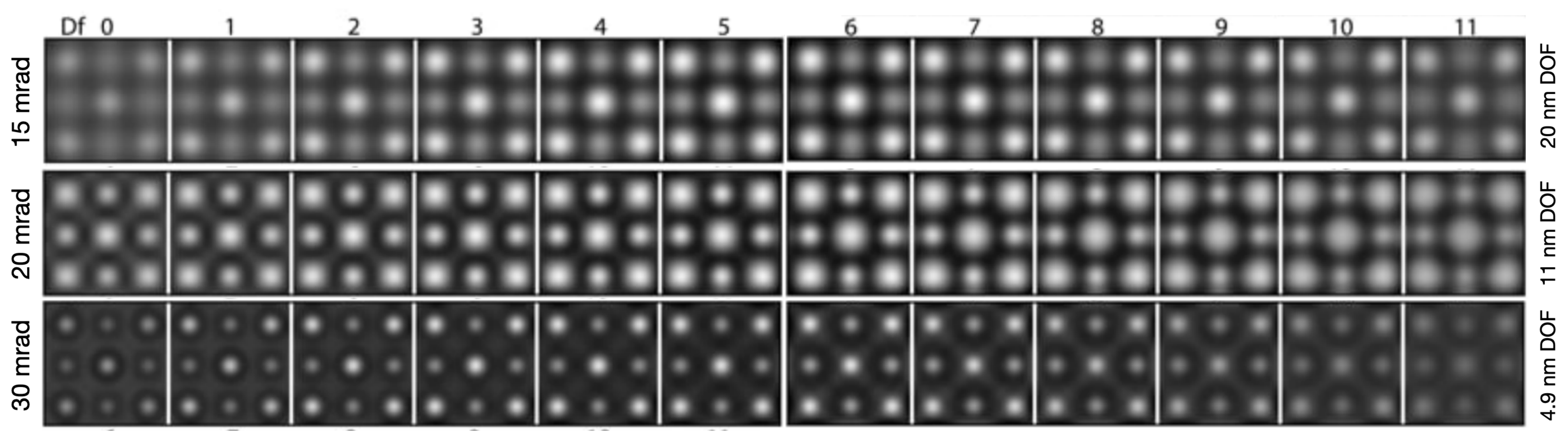}%
 \caption{\label{12nm_dfvsca} { SSB images of very 12 nm thick STO as a function of physical probe focus within the sample depth (zero defocus (df) is again the entrance surface) as a function of the convergence angle. For this thickness the central focusing again seems to work regardless of the convergence angle.}}
\end{figure}

At 200 kV the 20 mrad convergence angle we used earlier, the depth of field is 11 nm. So one might think that 5 nm of STO, with less sample potential within the depth of field, would show less contrast reversals than 16 nm or more of STO with the same conditions. Interestingly, this is not what is observed. To further explore this we turned to 12 nm thick STO simulations. As shown in figure \ref{12nm_dfvsca}, regardless of if the convergence angle produces a depth of field similar to, much greater than or much less than the 11 nm thickness of the material, central focusing is again optimal.

Dynamical effects are generally more associated with thicker samples, and 5 nm is certainly on the thin side, so one would expect dynamical scattering to effect the contrast of samples thicker than 5 nm more. Indeed, the divergence from central focusing being optimal seen at 50 nm of STO, where focusing to the exit surface was instead found to be optimal, seems more naturally a result of strong dynamical effects being balanced by defocus. 
The convergence angle of course weights the frequencies differently, as is already usually considered when optimizing the contrast transfer function without considering contrast reversals. However different frequency weightings will also naturally change where asymptotes occur in the frequency response, due to phase wrapping in probe reciprocal space, which in turn effects the appearance of contrast reversals.  


Given that optimal contrast is generally obtained for ADF with the probe focused to the entrance surface and generally for both iCoM and direct ptychography methods with the probe deeper within the sample, the optimal focus for ADF and iCoM and direct ptychography methods such as SSB can increasingly deviate as the sample increases in thickness. Different strategies can therefore be employed depending on the needs of the sample. One can optimize for the iCoM and ptychography by aiming to put the probe deep within the sample at the cost of contrast in the ADF signal, or one can put the focus to the entrance surface as is best for the ADF. In practice the first option has been easier to achieve experimentally, using the ADF to guide selecting a region of interest and final tuning, including the focus. 

However live processing options becoming available for both iCoM \cite{real_time_icom} and SSB \cite{Live_SSB1,Live_SSB2} are beginning to enable the focus to be optimized for these imaging modes while scanning. It is nonetheless important to remember that the ptychographic contrast can also be adjusted significantly post collection \cite{Chuang_APL} either by post collection defocus or other schemes to adjust the phases. The choice of focusing strategy will depend on the dose the sample can handle and the importance of the Z-contrast ADF signal, which remains very useful for correct interpretation when sufficient Z-contrast signal is available. The importance of Z-contrast is all the more so when contrast reversals can occur with phase imaging. Thus it is very useful to understand the overall contrast variation of direct ptychography methods with defocus and thickness as well as the convergence angle.

\section{Conclusion}
We have provided an explanation of why the overall strength of direct SSB electron ptychography is often at its maximum when the probe is focused midway through the sample. By focusing the probe to the center of a sample, equal and opposite defocus induced phase cancels out at sample planes equally far from the focal plane of the probe. This removes most of the defocus contribution from samples planes away from the focal point of the probe allowing samples to be seen most clearly when other effects such dynamical scattering are not overly strong. We demonstrated this using the iCTF concept in combination with simulations of STO, but the concept can be expected to extend to many materials typical of materials science. In addition we have demonstrated the convergence angle can be a crucial factor in determining if contrast reversals will be present at all and if central focusing will remove them. This knowledge will therefore help inform the design of focused probe ptychography experiments of thicker samples in determining where exactly to focus the probe.

\bibliography{chuang2refs}


\end{document}